\documentclass[a4paper,11pt]{article}
\pdfoutput=1 

\usepackage{jcappub} 

\usepackage[T1]{fontenc} 

\usepackage{epsf,amsmath,bbold,amsfonts,stmaryrd}
\usepackage{mathrsfs}
\usepackage{appendix}
\usepackage{amssymb}
\usepackage{amsthm}
\usepackage{float}
\usepackage{color} 
\usepackage{color}
\usepackage{lineno,hyperref}
\usepackage{graphicx}
\usepackage{dcolumn}
\usepackage{bm}
\usepackage{dsfont}
\usepackage[english]{babel}
\usepackage{empheq}
\usepackage[theorems,skins]{tcolorbox}
\usepackage{caption}
\usepackage{url}
\usepackage{enumerate}
\usepackage{slashed}
\usepackage{verbatim}
\usepackage{esint}
\usepackage{bigints}
\usepackage[capitalise]{cleveref}
\usepackage{upgreek}

\title{Small-scale Cosmic Signatures of Feebly Interacting Massive Particles}

\newcounter{PHQ}

\author[a,b]{Patrick Hager}
\author[b]{and Alexis Kassiteridis}

\affiliation[a]{Physik Department T31, 
	Technische Universit\"at M\"unchen,\\ 
	James-Franck-Stra{\ss}e 1, 85748 Garching, Germany}
\affiliation[b]{Arnold Sommerfeld Centre for Theoretical Physics,\\ 
	Theresienstr. 37, 80333 M\"unchen, Germany}

\emailAdd{patrick.hager@tum.de}
\emailAdd{a.kassiteridis@physik.uni-muenchen.de}

\abstract{Feebly Interacting Massive Particles (FIMPs), if they exist, should be notoriously difficult to detect even indirectly. In order to constrain them, 
	we derive bounds for feeble theories sourced via Standard Model fields by investigating their predicted signatures regarding small-scale structure formation.
	To achieve this, we obtain an analytic approximation for the phase-space distribution function for a generic dimension $\ell$ scattering operator.
	As a proof of concept, realizations of such theories are discussed, which provide a viable thermal evolution and are able to solidly solve the enduring small-scale structure challenges appearing in $\Lambda$CDM cosmology.}


\arxivnumber{2009.11308}

\begin{document}
\maketitle
\flushbottom

\section{\label{sec:level1}Introduction}

Until now, no direct nor indirect detection of dark matter (DM) modes has been observed, although various signals hint towards their particle nature. 
The lack of clear experimental evidence may possibly hint to the existence of  feebly interacting dark matter (FIMP) candidates as DM. 
The main aim of this work is to investigate theories of FIMPs sourced directly or indirectly by the Standard Model fields by computing their effects on small-scale structure formation. These theories could provide solutions to the enduring $\Lambda$CDM issues.

Such theoretical scenarios are subject to several constraints, enlarging the predictive power of the allowed theories. 
Therefore, we are interested in theories of FIMPs extending the Standard Model of particle physics (SM) minimally; feeble in the sense of the extremely weak and out-of-equilibrium interactions between the FIMPs and the SM modes.
We achieve our goal by studying these theories at the level of the phase-space distribution function of the participating FIMPs. Here, we compute an analytic approximation of the distribution function for generic effective contact interactions.


At the end of the 20th century, the sole purpose of a DM candidate was to solve the large-scale structure problems. However, critical questions regarding the cosmic structure formation at small scales have arisen. Moreover,  within $\Lambda$CDM, there are four main small-scale problems concerning DM, for a pedagogical introduction see e.g. \cite{Bringmann:2016ilk}. More precisely, recent observations are found to be in tension with the predictions of CDM simulations at small scales \cite{Vogelsberger:2014pda}; this is known as “the small scale crisis of $\Lambda$CDM” and it mainly consists of the cusp vs. core problem and the too-big-to-fail problem, together with the missing satellite and diversity problems. 

The cusp vs. core problem describes the clear discrepancy between the cusp-profiles of  the DM density emerging from high-resolution CDM simulations \cite{Hut:1984qd}, where the halo densities follow the Navarro-Frenk-White profile \cite{Tegmark:2003uf}, and the observed rotation curves of disk galaxies, which favour a cored, less steep inner density profile \cite{Bahcall:1999xn}.  Another small scale problem is incarnated via the following puzzle: although the satellite galaxies of the Milky Way exhibit the largest stellar velocities, and therefore should be part of the most massive subhalos of the Milky Way, the CDM simulations predict that these subhalos must be too dense in their central regions to be suitable for hosting satellites according to observations \cite{Bertone:2004pz}. In other words, although these satellites belong to the most massive ones, they are predicted to be poor in visible matter and to suppress the formation of galaxies; namely, these cosmic subhalos are too big to fail to form stars (luminous matter). Furthermore, the CDM simulations predict a large number (order of 100 or even 1000) of subhalos inside the CDM halos \cite{Cheng:2002ej}, however, only a fraction of such halos had been discovered, when the missing satellite problem emerged \cite{Preskill:1982cy}. Finally,  as an extension of the cusp vs. core problem, the cosmic structure formation should lead to little scatter in profiles of halos with given mass \cite{Goldberg:1983nd} according to CDM simulations, however this is not the case for the observed disk galaxies \cite{Servant:2002aq}, which admit large variation of central densities; this issue is known as the diversity problem of $\Lambda$CDM.

 While it is possible that these small-scale problems are due to baryonic physics or modifications of gravity \cite{Tulin:2017ara}, we entertain the idea that these issues might be fixed within the dark sector itself. More preciscely, the solution to all these problems is quite straightforward after the correct choice of the parameter space in a self-interacting dark matter theory. In order to illuminate the underlying mechanism, the following schematic explanation is given: inside a dense halo center, the particles with low entropy are being consecutively heated via self-interactions and especially via 2-to-2 scatterings. This process results in a reduction of the inner density (i.e. cusp becomes core), as heated particles move to outer regions of the halo. In addition, such self-scatterings also affect the substructure of halos, leading to a possible solution of the missing satellite problem \cite{Borriello:2000rv}. It is shown that SIDM is able to solve the too-big-to-fail problem \cite{Zavala:2009ms} and diversity problem \cite{Lovell:2011rd} as well.

In this work, we use the distribution function to constrain FIMPs from their corresponding signatures of the small-scale structures, as opposed to the large-scale constraints \cite{Heeck:2017xbu, Boulebnane:2017fxw,Kamada:2019kpe}. Furthermore, we study feeble theories and find solutions of the small-scale problems. In other words, we consider these small-scale problems as further constraints on top of the large-scale constraints that are already well-understood.
Hence, our work focuses mostly on understanding how these small-scale challenges are met, and provide solutions in the context of feeble theories for the small-scale structure problems of $\Lambda$CDM.
Although these bounds are much tighter than the ones inferred by collider measurements, they lead to acceptable theories alleviating the small-scale crisis of $\Lambda$CDM cosmology. 
In particular, we show that recent Lyman-$\alpha$ bounds  \cite{Irsic:2017ixq}, together with the positive observations regarding the rotation curves, the missing dwarf satellites, and the cored profiles of galaxies discussed in \cite{Vogelsberger:2015gpr, Schewtschenko:2015rno, Tulin:2017ara}, indicate the presence of new physics well below the weak scale.
This secluded dark sector turns out to be strongly interacting with itself, departing from the common idea of weakly interacting particles (WIMPs) and FIMPs \cite{Tulin:2017ara}.

The history of feeble interactions accommodated in dark matter models starts long ago. For example, Higgs-sourced feeble models acquired attention after \cite{McDonald:2001vt}; however, the list of the corresponding constraints tends to not include all relevant ones inferred from the small-scale structure measurements. 
For a thorough review of such theories, see \cite{Bernal:2017kxu} and \cite{Matsui:2015maa}. 
The treatment of these theories on the level of distribution functions has recently gained some attention \cite{Bae:2017dpt, Dvorkin:2019zdi}, and we extend this treatment by giving analytical approximations for general dimension-$\ell$ interactions between the SM and the dark sector. We give more details on this in \cref{sec:2}.

The main aim of our work is to show that a family of SM sourced theories enables a minimalistic scenario of viable thermal evolution, solving the enduring small-scale challenges appearing in $\Lambda$CDM cosmology. The proposed generic theories are characterized  by three checkpoints in the thermal evolution; in other words, three temperatures $T_0, \, T_1,\, T_2$, the reheating, the production, and the rest-mass temperatures respectively, are responsible for the resulting cosmic structure at large and small scales. To the best of our knowledge, our results, and our mass intervals, yield, for the first time, much more concrete predictions about the minimal DM extensions of the SM than found in \cite{McDonald:2001vt, Hardy:2018bph, Heeck:2017xbu, Boulebnane:2017fxw, Kamada:2019kpe}; in a sense, we concentrate more on confronting the small-scale issues of $\Lambda$CDM rather than the general model building of feeble theories. 
Furthermore, we claim that the theoretical prototypes presented in this work satisfy all experimental and theoretical constraints, and at the same time provide clear solutions to the missing satellite and high values of rotation curves problems, together with the cuspy simulated profiles and the diversity problem. 

Our approach, although respecting the same constraints and solving the same problems as the minimal flavored dark QED theory \cite{Balducci:2018ryj} or the neutrinophilic framework presented in \cite{Balducci:2017vwg}, is very different, since the DM modes are the FIMPs sourced entirely from the Higgs field.
On the other hand, the feeble theories could, in principle, admit also a leptophilic or neutronphilic virtue, discussed in \cite{Balducci:2018dms} and in \cite{Karananas:2018goc}, explaining the long standing neutron anomaly, respectively.

Recently, the XENON collaboration in the XENON1T experiment observed an excess in the electron recoil energy in the region $E\lesssim 7 \rm keV$ compared to the known background \cite{Aprile:2020tmw}. This is a strong motivation, independent from the small-scale problems, to investigate the possibility of dark matter which admits masses of a few keV. Therefore, in this work, we will study feebly theories, which accommodate modes with masses at the keV regime.

This paper is organized as follows. 
We start by deriving the general phase-space distribution function of the feeble particles for generic dimension $\ell$ operators. Then, we continue with a sufficient approximation for the cut-off scales of the linear power spectrum generated by the feebly interacting particles at the level of the phase-space distribution function, and we present our list of constraints arising from the study of the small-scale structures in theories of FIMPs, sourced directly or indirectly by the Standard Model fields. 
As a pedagogical example, we start with DM productions via Higgs decay, and we generalize this procedure to study more complex production mechanisms.
We conclude by showing that, although the allowed spectra are tight, the existence of different paradigms as prototype classes of such theories could be still possible, and could provide viable solutions to the small-scale challenges of $\Lambda$CDM cosmology.

\section{\label{sec:2}The General Phase-space Distribution Function}

In this section, we present an analytical solution of the Boltzmann equation for the general SM-sourced case.
We consider a generic interaction term
\begin{equation}
\mathcal{L}_{\rm int} =  g \Lambda^{4-\ell}\, \Phi_{\rm SM}... \Phi_{\rm SM}\, \Phi_{\rm DM}... \Phi_{\rm DM}+ \mathcal{L}_{\rm self}[\Phi_{\rm DM}] \label{interaction vertex}
\end{equation} 
in the Lagrangian density, where $\Phi_{\rm SM}$ denotes  any SM field, and $\Phi_{\rm DM}$ denotes the DM fields, $g$ is a coupling and $\Lambda$ is the energy suppression of the corresponding operator. The first part of the Lagrangian density is the DM-SM portal, and the second one  encapsulates possible self-interactions inside the dark sector.

The defining equation of the DM distribution function $f(\textbf{p}(t))$ per degree of freedom (d.o.f.) of the underlying field, in a Friedmann-Robertson-Walker universe with Hubble function $H$, yields
\begin{equation}
(L-C)[f](\textbf{p},t)=0\; ,
\end{equation}
where $L:= p^0(\partial_t-H\textbf{p}\cdot \nabla_{\textbf{p}})$ is the Liouville operator, which gives the change of $f$ with respect to an affine parameter along a geodesic, $H$ is the Hubble function; $C$ is simply the collision term encapsulating the probability of this change, and, generically assuming CP invariance of the $S$-matrix, is given by
\begin{equation}
\begin{split}
C[f](\textbf{p}):=\tfrac{1}{2 V^{(4)}}\sum_{\rm in,out}&\left( \prod_{i\neq \{\textbf{p}\}} \int {\mathrm{d}\omega_i}\right) \vert \langle {\rm out} \vert (S-1) \vert{ \rm in} \rangle\vert^2\\
&\cdot \left[ \prod_{j\neq \{\textbf{p}\}} f_{\rm out} \prod (1\pm f_{\rm in} ) - \prod f_{\rm in} \prod_{j\neq \{\textbf{p}\}} (1\pm 	f_{\rm out} )\right]\; ,
\end{split}
\end{equation}
with $V^{(4)}$ the appropriate space-time normalization volume for all d.o.f.\ of the underlying field. 
We used the abbreviation in/out for the initial/final, identically free states, and $ {\mathrm{d}\omega_i}$ for the Lorentz-invariant measure; the $\pm$ stands for Bose-Einstein-/Fermi-statistics respectively. 

Direct integration has previously been used to great success to solve for the FIMP distribution function numerically \cite{Bae:2017dpt,Dvorkin:2019zdi}. However, we proceed in a different manner, and solve for the distribution function in terms of the comoving momentum, leading to an analytical expression that is vital for the straightforward treatment of the linear power spectrum cut-off.

It is convenient to write the Liouville operator in terms of the comoving momentum ${\uppi \equiv  a(t)\, p\vert_t/a(t(M))M}$ \cite{Bringmann:2006mu} at some reference temperature $M$, introducing the reduced source-mass $\mu \equiv \Lambda/T$. The operator then takes the form
\begin{equation}
L[f](\uppi,\mu) = \frac{3 p^0 H}{\frac{3}{\mu}- \partial_{\mu} \log [g_*^s(\mu)]} \partial_\mu f (\uppi,\mu)\, .
\end{equation}
Here, $\partial_\mu$ is the partial derivative with respect to the reduced source mass. 
In other words, it is straightforward to start our approach by modelling the DM distribution function $f(\uppi,\mu)$ as a solution to the Boltzmann equation; this way allows us to obtain all relevant observables directly by integration of the now known phase-space distribution function.


We start by assuming that the relativistic degrees of freedom vary slowly with respect to the temperature, $\partial_\mu g_*^{(s)}(\mu) \approx 0$, during the DM production epoch, which is after the electroweak phase-transition and before the bottom quark annihilation. 
Without loss of generality, we take 
\begin{equation}
\uppi = \left.\frac{p}{T}\right\rvert_{t_1}
\end{equation} 
for convenience, where $t_1$ is the time of the DM production. 

As an example, we consider a minimal extension of the Standard Model, where only a single Higgs field couples to pair of DM scalars, setting $\Lambda =m_{\rm Higgs}$ equal the Higgs mass. 
Namely, we consider the FIMP production only via $1\rightarrow 2$ Higgs decay into a DM pair, taking place after the electroweak phase transition. This is due the strong suppression of the Higgs annihilations into DM after the phase transition for the given parameter space of masses and couplings, since the massive Higgs modes follow their equilibrium abundance. After some algebra and numerics, we determine the form of the DM distribution function $f(\uppi, \mu)$ per feeble mode.
For the Higgs decay into two feeble particles, where the Higgs boson is in local thermal equilibrium (LTE), we obtain
\begin{equation}
f_{\mathrm{DM}}^{1 \rightarrow 2}(\uppi,\mu) = f_0(\mu) \frac{\mathrm{e}^{-\uppi}}{\sqrt{\uppi}} \left(\,  \text{erf}\left(\frac{\mu}{2 \sqrt{\uppi}}\right)-  \frac{\mu}{\sqrt{\pi\uppi}}\,   \mathrm{e}^{-\mu^2/4 \uppi}\right)\, . 
\end{equation} 
Here, $f_0(\mu) $ is the model-dependent amplitude, which varies very slowly as $\mu$ changes, we choose the reference temperature as the production one $M=T_1$, as before, and $\pi$ is just the real number $\pi$. 

One notices that the feeble distribution function is almost thermal at large values of $\mu$, admitting a first moment of $\langle \uppi \rangle \approx 2.45$. Moreover, the phase-space distribution function reduces to
\begin{equation}\label{fDM Approx}
f_{\mathrm{DM}}^{1 \rightarrow 2}(\uppi) \approx \lim_{\mu\rightarrow \infty} f(\uppi, \mu) = f_0\, \frac{ \mathrm{e}^{-\uppi}}{\sqrt{\uppi}}
\end{equation} 
at later times, which trivially solves the collisionless Boltzmann equation, encapsulating the notion of feeble interactions between the produced particles and the Higgs boson.
It turns out to be a good first approximation of studied observables, since the coefficient of variation (CV) is $\sim 0.6$, the same as in the case of ultra-relativistic fermions or bosons in LTE.  
We note that the DM is populated as long as the hot DM bound of 1$\%$ \cite{Ade:2015xua} is achieved, namely around $\mu=9.1\equiv \mu_1$; this is the proper definition of $t_1$: the time when the DM relic abundance has been populated.

The above result is obviously process-dependent. Therefore, we take the effort to compute, for the first time, the DM distribution function in the general case of SM-sourced FIMPs before the electroweak phase transition, where a pair of SM modes annihilates into DM via a dimension $\ell$ operator of massless modes.
This production mechanism  is actually the high energy limit of the scenario studied above: the SM modes are still ultra-relativistic and produce DM modes via feeble interactions.
Without loss of generality, we consider the case of $2 \rightarrow 2$ processes; technically, since it is the process which is the least suppressed in the phase space. Furthermore, the source modes are assumed to achieve LTE at a temperature $T_0<\Lambda$, which could be the reheating temperature.
We obtain the analytic solution
\begin{equation}\label{phase space 2to2}
f_{\mathrm{DM}}^{\ell ,\, 2\rightarrow 2}(\uppi, \mu) =f_0(\mu,\mu_0)\, \frac{\mathrm{e}^{-\uppi} }{\uppi^{5-\ell}}\, .
\end{equation} 
%
Again, $f_0(\mu,\mu_0)$ varies very slowly with respect to $\mu$ and depends polynomially on $\mu_0=T_0/\Lambda$ absorbing $g$ in $\Lambda$, which makes the DM distribution function independent of $\mu$ at the lowest order. For example, this describes the possible mechanisms where high energy muons could scatter and create DM relics, which might be interesting, since such processes contribute to the anomalous magnetic moment of the muon.\\

Although feeble interactions are extremely weak, self couplings inside the dark sector are not necessarily suppressed. At least after a given point in time during the evolution of the universe, such interactions could freeze-in, leading to a thermalization of the DM population. This event can change the macroscopic properties of the relics, for example, the particle density, the evolution of DM fluctuations, and even the phase-space distribution function. For self-scattering processes of the lowest order (2 $\rightarrow$ 2), the integrated expression of the collision term is proportional to
\begin{equation}
\left( \prod_{i=\{1,...,4\}} \int {\mathrm{d}\omega_i}\right) \vert \mathcal{A}_{\rm self}\vert^2 \, \delta^{(4)}\left(\sum p_i\right)\,  \left[  f_{3} f_{4}  -  f_{1} f_{2} \right]\, ,
\end{equation}
using the shortcut notation $f_i \equiv f(\textbf{p}_i(t))$, and $\mathcal{A}_{\rm self}$ denoting the self-scattering amplitude. 
This collision term gives the change of the DM number of particle 1, which is conserved due to symmetry. 
This is of great importance, since it implies that the DM relic abundance can not be modified via self-interactions at the lowest interaction-level; higher interaction-levels are required, which are much more suppressed. 
Furthermore, one could exemplary examine perturbations of the peculiar velocity of the DM fluid \cite{Bringmann:2006mu} in the context of self-interactions as in \cite{Bringmann:2016ilk}, given by
\begin{equation}
\left( \prod_{i=\{1,...,4\}} \int {\mathrm{d}\omega_i}\right) \vert \mathcal{A}_{\rm self}\vert^2 \, \delta^{(4)}\left(\sum p_i\right)\, \frac{\textbf{n}\cdot \textbf{p}_1}{m}  \left[  f_{3} f_{4}  -  f_{1} f_{2} \right]\,,  \label{self-int zero}
\end{equation}
with $\textbf{n}$ the wave vector of the linear decoupled perturbation Fourier modes in a spatially flat background \cite{Green:2005fa}. 
Here, we assumed that the DM modes with mass $m$ are non-relativistic, taking $p^0_i \approx m$. 
One notices that also this collision integral vanishes due to  the zero contributions of the $\textbf{p}_1\pm\textbf{p}_2$ moments. 
This means that the lowest order of the self-interactions of non-relativistic DM particles does not contribute to the evolution of DM fluctuations \cite{Bringmann:2016ilk} allowing us to use the results of \cite{Bringmann:2006mu} even in the context of a theory with self-interactions.\\
In order that the DM distribution function exhibits significant deviations away from our analytic solutions, \cref{fDM Approx,phase space 2to2}, higher order processes should first freeze-in; namely, scattering channels that are able to change the population of DM such as $2\leftrightarrow 4$. One finds in the case of scalar FIMPs that the change of the distribution function after the freeze in of self-interactions is of order $\mathcal{O}(\lambda^4 f_0^2)$, where $\lambda$ is the effective self-coupling in the dark sector. Similar problematic also applies in the case of fermionic DM. The consideration of these corrections do not change the results of this work in the keV regime, as we will explain in more detail when we study the large-scale cosmic structure.

\section{Linear Power Spectrum Cut-off}

One of the main purposes of this work is to examine whether feeble theories could solve all the small-scale problems of $\Lambda$CDM, which at the same constrain the FIMPs due to present bounds arising from the observed small-scale structures.
In this section, we show how feeble theories could in fact admit interesting signatures at small scales.

\subsection{Efficiency of Momentum Exchange}

We start by reviewing the general assumption in WIMP theories, stated in \cite{Balducci:2018ryj} and \cite{Balducci:2018dms}, that the cold DM modes are in local kinetic equilibrium with some abundant  (usually ultra-relativistic) species. 
This mechanism is described by the efficient momentum exchange between the participating particles; however, as described in \cite{Bringmann:2016ilk}, the elastic scattering processes between the DM modes and the plasma lose their efficiency at times around the kinetic decoupling temperature $T_1$. 
In \cite{Hofmann:2001bi}, it is thoroughly explained that the remaining elastic interactions can  be described effectively as sources of entropy production in an imperfect DM fluid; in other words, the momentum exchange continues to damp the perturbations that would otherwise grow to form the first gravitational bound DM objects (protohalos) \cite{Green:2005fa}. 
Therefore, the DM particles stream free of damping and efficiently erase the small-scale structures. 
This entropy production is found to be related to the dissipative evolution of DM fluctuations, which are exponentially damped with a characteristic mass scale $M_{\rm d}$; in other words, small-scale structures of DM cannot sustain their form within the Hubble volume at the time of kinetic decoupling, since the plasma-particles within this volume are much more abundant and their interactions suffice to keep the former in approximate local thermal equilibrium. Furthermore, this damping mass scales with $T_1^{-3}$ \cite{Balducci:2018ryj}, the exact relation is given in \cite{Vogelsberger:2015gpr} and we cite it here,
\begin{eqnarray}
M_{\rm d}
\approx
7\times 10^{10} \; 
\left(\frac{0.1\, \rm keV}{T_1}\right)^3 \; M_\odot
\; ,
\end{eqnarray}	
where $T_1$ is SM-photon temperature. 
One notices that the above expression is independent of the mass of the DM particles, since at the time of kinetic decoupling they are almost at rest (cold). 
It is important to note that this problematic is valid only if the DM modes are in local kinetic equilibrium with the abundant plasma, and, therefore, it is not present in a theory with feeble interactions of a single DM species, even if the DM particles are in local thermal equilibrium with themselves via self-interactions. 
This is due to the symmetries of the collision term describing the self-scattering between the DM particles. 

Summarizing the previous results and recalling  that the lowest interaction-level of the self-interactions of non-relativistic DM particles, \cref{self-int zero}, does not contribute to the evolution of DM fluctuations, in a feeble theory of a single DM species, no notion of kinetic decoupling is present, and self-interactions do not allow us to extract information about the cut-off masses of the linear matter power spectrum. 
Therefore, we search for a different damping mechanism: the free-streaming of the feeble modes.

\subsection{The Free-streaming of Feeble Modes}

When a particle at time $t_1$ starts moving freely through the expanding universe, its free-streaming length also provides a different mechanism leading to a cut-off of the linear matter power spectrum \cite{Green:2005fa}. 
The length of the free propagation of the DM modes after the $t_1$-surface, moving freely along geodesics, up to the matter-radiation equality time, $t_{\rm{eq}}$, is defined as
\begin{equation}
\ell_{\rm fs}=   \int_{t_1}^{t_{\rm{eq}}}\mathrm{d} t\, \frac{v(t)}{a(t)}\, .
\end{equation}
The approximate limit until $t_{\rm{eq}}$ is due to the rapid structure formation after that time. Moreover, the corresponding cut-off mass is estimated after taking  $\ell/2$ as the radius of the underlying homogeneous matter sphere.

For cold particles that are already non-relativistic at the time of kinetic decoupling, if such regime is present, the cut-off mass yields \cite{Green:2005fa}
\begin{equation}
M_{\rm fs} \approx 4.2 \times 10^{8} \left(\frac{1+ \log \left[g^{1/4}_*(T_1)\, \left(T_1/{5\, \text{keV}}\right)\right]/10.2}{\left({m}/{100\, \text{keV}}\right)^{1/2} g^{1/4}_*(T_1)\, \left(T_1/{5\, \text{keV}}\right)^{1/2}}\right)^3  \; M_\odot \, .
\end{equation}

However, in this work we are interested in FIMP production mechanisms sourced by the Higgs field; therefore, the DM modes are born relativistic mostly at $t_1$, even if their mass is much larger than the plasma temperature. 
This surface coincides with the production time of the particles, but not with the time $t_2>t_1$, when the DM modes become non-relativistic. 
The free-streaming length of such modes can be approximated as
\begin{equation}
\ell_{\rm fs} \approx 2 \frac{z_{\rm eq}^2}{z_2} t_{\rm eq} \left[1+ \log\left[\frac{z_2}{z_{\rm eq}}\right] + \mathcal{O}\left(z_2 \frac{T_0}{m_e}\right) \right] \, , 
\end{equation}
assuming that $t_2\gg M_{\rm{Pl}}/m_e^2$, and 0 denotes the observable values at present time. 
We also considered the change of the expansion parameter $a(t)$ as a function of time, but such changes do not affect the above result up to order of $z_2 T_0/m_e$. 
Here, plugging in the most recent results inferred from Planck measurements \cite{Ade:2015xua}, one obtains
\begin{equation}\label{free stream}
\ell_{\rm fs} \approx 0.048\, \text{Mpc}\,  \left(\frac{10^{7}}{z_2}\right) \left[1+ \log\left[\frac{z_2}{3365}\right]\right]\, ,
\end{equation}
with the redshift at the time where the DM modes enter the non-relativistic regime denoted by $z_2$, given by 
\begin{equation}
z_2={\uppi_1}^{-1}\frac{m}{T_*},
\end{equation}
assuming a $1\rightarrow2$ decay channel. 
Here, $m$ is the DM particle mass and $\uppi_1$ is the comoving momentum at the time of production, taking $M=T_*$. 
Exemplary, if the FIMPs become non-relativistic at $z_2 \sim 10^7$, they  admit warm dark matter (WDM) properties regarding the small-scale structure formation \cite{Vogelsberger:2015gpr}. 
The corresponding numerical estimate bounds the DM mass stronger from below than the naive entropy dilution calculation \cite{Matsui:2015maa}. 
In a FIMPly theory, the cut-off mass to the linear matter power spectrum is the smallest possible protohalo \cite{Bringmann:2006mu} that could be formed, and is given by $M_{\rm ph}\equiv M_{\rm fs}$, since the damping due to acoustic oscillations is absent. 
However, this does not encapsulate the change of the phase-space distribution function of the DM particles over their production time. 
Hence, we will now compute the free-streaming length directly from $f(\uppi,\mu)$.
The cut-off of the linear power spectrum is estimated via
\begin{equation}\label{mass approx2}
M_{\rm{fs}}(\mu_1)= N(\mu_1)^{-1}\int_0^\infty \mathrm{d}\uppi\, \uppi^2\,  M_{\rm{fs}}(\uppi)\, f(\uppi, \mu_1)\, ,
\end{equation}
where $N(\mu)$ is the appropriate temperature dependent normalization of the distribution function, independent of the reference temperature $M$.
%


\section{Cosmic Structure due to Feeble Theories}

At this point, we use the phase-space distribution function of the DM particles in order to study feeble theories under the spell of cosmological observations at large and small scales. Moreover, we invert present constraints and we show that such theories admit a parameter space which is large enough to be able to solve the enduring small-scale problems appearing in the $\Lambda$CDM paradigm, while producing successfully the same large-scale cosmic structures. To do so, we discuss a generic way towards a minimal model of self-interacting FIMPs, such that our approximation of the distribution function can be used. 

\subsection{Feeble Theories at Dwarf Scales}

In this section, we show that the allowed mass spectrum of the feeble theories sourced by SM fields alleviates the missing satellite problem, respecting all present constraints. More precisely, it is suggested by the authors of \cite{Vogelsberger:2015gpr}, \cite{Boehm:2000gq} and \cite{Aarssen:2012fx}, that an exponential suppression of the linear matter power spectrum at dwarf scales (sub-kpc distances) alleviates the missing satellite problem of $\Lambda$CDM cosmology. 
The sufficient cut-off mass should be of order of magnitude  $10^8$ solar masses and not above $10^9$ solar masses; such protohalo masses alleviate the small-scale abundance problem \cite{Bringmann:2016ilk, Tulin:2017ara}.

If the proposed DM theory possesses a kinetic decoupling regime with the DM modes already being non-relativistic, a kinetic decoupling temperature $T_{\rm 1} \sim $ keV  provides such cut-off masses \cite{Tulin:2017ara}. 
On the other hand, as explained in the last section, in DM theories of feebly interacting particles the above regime is absent and the protohalo masses are mainly determined through the free-streaming mechanism and correspond to the WDM mass.
These masses are in turn constrained by the Lyman-$\alpha$ bound, as explained in \cite{Aarssen:2012fx, Bullock:2010uy, Baur:2015jsy}, and more recently in \cite{Irsic:2017ixq}; in other words, the maximum cut-off mass almost touches $10^9 M_\odot$, or, equivalently, a minimum of 3.5 keV warm DM mass \cite{Bringmann:2016ilk} for the weak bound, and lies somewhat above $10^8 M_\odot$, which corresponds to a 5.3 keV warm DM mass, for the strong bound. Even values around $7-8$ keV due to the form of \cref{mass approx2} can still marginally alleviate the missing satellite problem, and are even in line with the recent study on warm DM mass bounds \cite{Kamada:2019kpe}; however, the authors considered production mechanisms which differ from the purely SM sourced channels that we study, namely included new heavy modes, which perplex the production amplitudes and therefore their derived bounds a not directly applicable on our work.

\begin{figure}
	\includegraphics[scale=0.6]{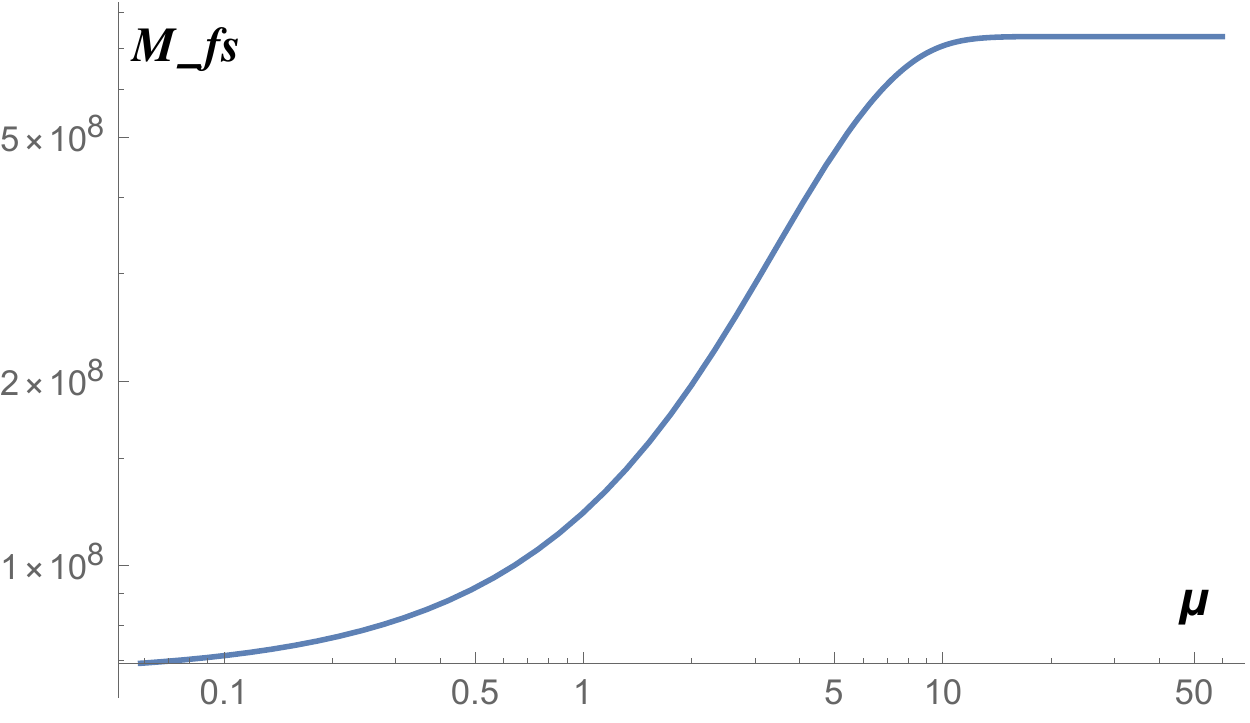}
	\caption{The cut-off mass due to the free-streaming of the DM modes with mass $m=3.55$ keV as a function of the reduced mass of the decaying source particle. One notices that at $t_1$ the cut-off mass admits almost its maximum value.}
	\label{free-streaming}
\end{figure}

This problematic, together with the most recent Lyman-$\alpha$ constraints and the thermal approximation \cref{fDM Approx}, using the DM phase-space distribution function, yields a lower bound on the DM mass after assuming $1\rightarrow2$ decay processes of the Higgs boson and fixed DM relic density $\Omega_{\rm{DM}}h^2=0.12$, namely, \cref{mass approx2} yields
$
m \gtrsim 5.8\, (3.1) \, \mathrm{keV},
$
regarding the strong (weak) constraint. Almost the same bounds appear if $2 \rightarrow 2$ production ways before the electroweak phase transition are considered as the main production mechanism. Such values are very close to the ones found in the literature regarding the lowest warm DM mass \cite{Irsic:2017ixq}. This result, although process-dependent, due to the form of $f_0$, turns out to be a very good first approximation for $M_{\rm{ph}}$, since the coefficient of variation (CV) is $\sim 0.6$, the same as in the case of ultra-relativistic fermions or bosons in LTE. 
We find that if the DM modes are stable, then the limit $\lim_{\mu\rightarrow \infty}  M_{\rm{fs}}(\mu)$ exists, and is only about 5$\%$ larger than the production estimate at $\mu_1$, as plotted in \cref{free-streaming}.
If one uses the warm DM distribution due to non-relativistic decays \cite{Kamada:2013sh,Strigari:2006jf}, the free-streaming length is smaller by 10$\%$ due to the non-constant momentum distribution.
A more accurate bound is beyond the scope of this work, since we are interested in showing that the allowed mass spectrum of the feeble theories sourced by SM fields alleviates the missing satellite problem, as proven before.

We note that it is also possible for heavier DM particles to generate similar cut-off masses, while being produced through feeble interactions, as long as their production takes place later than the typical decay time. 
This is, however, not possible if the decaying particle is the Higgs boson, since the feeble decay channel is not the dominant one. 
At least one additional portal field is needed in order to achieve $\mu_1 \gg 1$ as in \cite{Balducci:2017vwg}. 
Analogously, even lower DM masses are accessible if, at $\mu_1$, more relativistic degrees of freedom are present. 
Nevertheless, such modes should admit a similar temperature as the SM-modes, or be in pure LTE with them. 
Since no lighter-than-Higgs dark particles are observed, we assume the former case for the $1\rightarrow 2$ decay, which is strongly constrained during the big bang nucleosynthesis.

\subsection{Structures at Cluster Scales}

The defining property of feeble theories is the very weak interaction between different species, but this does not prevent strong self-interactions. 
Therefore, we turn our attention to the cosmological observations at cluster scales, and the arising constraints due to the existence of self-interacting DM (SIDM) in such theories. We show that the feeble theories sourced via SM fields admit a parameter space which provides possible solutions to the cusp vs. core, the too big to fail, and the diversity problems altogether.

The SIDM cross-sections produce effects that could significantly change the profiles and densities of massive subhalos; this happens during the non-linear structure formation. 
Furthermore, very large cross-sections would be incompatible with present measurements and very weak SIDM cross-sections are unable to solve the small-scale problems, mainly the cusp vs.\ cores and the too big to fail problem, together with the diversity problem; for a thorough discussion and review of all present (positive) constraints, see \cite{Tulin:2017ara}. 
For pure SIDM theories, where the damping of the matter power spectrum at the linear regime is of no importance,  dwarf galaxies with typical thermal velocities of order $\sim 10^{-4}$ for the  DM modes prefer large self interacting cross-sections,  $\langle\sigma_T/m\rangle_{v_{\rm therm}}\sim 1$ cm$^2$g$^{-1}$, as shown in \cite{Tulin:2013teo} and \cite{Tulin:2017ara}; however, at cluster scales with velocities of order $\sim 10^{-2}$, the constraints indicate that cross-sections larger than $\langle\sigma_T/m\rangle_{v_{\rm therm}}\sim 0.1$ cm$^2$g$^{-1}$ \cite{Tulin:2017ara, Elbert:2016dbb} are strongly disfavored, which could indicate a mild a velocity dependence of the SIDM cross-section between dwarf and cluster scales, similar to neutron-proton scattering. 

For example, assuming scalar DM, the cluster constraints give a clear lower coupling  of the effective quartic self-interaction $\lambda \Phi_{\rm DM}^4$,
\begin{equation}
\lambda \lesssim   10^{-4} \left(\frac{m}{50\, {\rm{keV}}}\right)^{3/2} \, .
\end{equation}
If one desires a velocity-dependent interaction for a more solid solution of the small-scale issues, the attention should be on a dark sector with some dark force mediator. For example, the corresponding bound of the effective Fermi coupling $\tilde{G}_F$ in a generic self-interaction term $\tilde{G}_F \left(\bar{\Psi}_{\rm DM}\Psi_{\rm DM}\right)^2$ is given by
\begin{equation}
\frac{\tilde{G}_F}{\mathrm{GeV}^{-2}} \gtrsim   4\times 10^{4}\,  \left(\frac{m}{50\, {\rm{keV}}}\right)^{-1/2}\, ,
\end{equation}
where $\Psi_{\rm DM}$ is a fermionic DM field.

Nevertheless, as studied in \cite{Schewtschenko:2015rno} and \cite{Vogelsberger:2015gpr}, the combined effects of the SIDM properties on the non-linear evolution at dwarf-galaxies, using significantly smaller SIDM cross-sections, and of the cut-off mass in late decoupling models (arising from a dark acoustic damping) provide simultaneous solutions to the enduring small-scale problems of the $\Lambda$CDM paradigm. 
It turns out that cut-off masses, which alleviate the missing satellite problem successfully, and SIDM cross-sections about an order of magnitude smaller than the ones proposed in \cite{Tulin:2017ara} solve the too big to fail, the diversity, and the cusp vs.\ core problem. 
Such a parameter set of SIDM with sufficiently large damping masses and milder SIDM interactions is provided by the ETHOS-4 tuned model \cite{Vogelsberger:2015gpr}: a bottom-up hybrid model, which is compatible with present cosmological constraints. 
This implies that higher values of the SIDM cross-sections tend to over-solve the $\Lambda$CDM issues. 

Hence, if the free-streaming cut-off masses are able to cumulatively affect the small-scale structure, then the FIMPs should admit SIDM virtues realized through some dark self-interaction; furthermore, the desired $\langle\sigma_T/m\rangle_{v_{\rm therm}}$ values at dwarf scales appearing in the ETHOS-4 tuned model \cite{Vogelsberger:2015gpr} are perfectly compatible with the above bounds of the effective quartic coupling for scalar DM, and the effective Fermi coupling for fermionic DM, respectively; this is due to the recent cluster results \cite{Elbert:2016dbb}. 
This problematic is also well motivated from the recent constraints on dwarf-scale cross-sections \cite{Bondarenko:2017rfu}, which are in tension with the proposed solution interval \cite{Tulin:2017ara}. 
However, these feeble models do not admit dark acoustic oscillations as ETHOS-4 tuned model, since no kinetic decoupling is present, meaning that further study and simulations are needed in order to examine the exact mapping of observables, which is beyond the scope of this work.

We note that both couplings are also constrained by perturbation theory and unitarity. This sets an upper bound on the DM mass of a scalar FIMP, hence
$
m \lesssim \mathcal{O}(10)\, \mathrm{MeV}.
$
Analogously, in the case of fermionic DM, the ETHOS-4 results and the cluster bounds yield multi-eV masses for the dark force mediators of the theory for multi-keV DM masses. 
This case may indicate the presence of new physics well below the weak scale.

\subsection{keV DM Masses in Big Bang Nucleosynthesis}

Now we track the footprints of feeble theories during the period of the big bang nucleosynthesis (BBN). 
To do so, we approximate the deviation of the effective neutrino degrees of freedom during that period. 
With the neutrinos already decoupled at $T_{\nu D} = 2.3$ MeV \cite{Enqvist:1991gx}, the corresponding deviation is given by the ratio 
\begin{equation}\label{delta effective}
\Delta N_{\rm eff} = N_\nu \frac{\rho_{\rm eff}}{\rho_\nu} \, ,
\end{equation}
with $\rho_{\rm eff}$ the energy density of the relativistic particle species besides the photons and neutrinos, with initial conditions being those at the neutrino decoupling temperature $T_{\nu D}$. 
This deviation works as a parametrization of the cosmic energy budget, and could affect the BBN processes significantly. Therefore, it can be probed at high precision. 

If $m\gtrsim T_{\nu D}$, then no change in $N_{\rm eff} $ can be seen. 
However, if $m< T_{\nu D}$, then at time $t$ and temperature $T_t$, we obtain
\begin{equation}
\frac{\left.\Delta N_{\rm eff}\right\rvert_t}{ N_\nu}  \approx  g_{\rm DM} \int \frac{{\rm{d}}^3p}{(2\pi)^3}\,  \epsilon(p)\, \left.f(p, T) \right\rvert_{t}/\rho_\nu \, ,
\end{equation}
with $g_{\rm DM}$ the degrees of freedom of the DM field and $\epsilon(p)=p^0-m$ the kinetic energy of the DM modes; their distribution is not thermal, as shown in \cref{sec:2}, therefore, one should ignore the rest energy of the particles. 
The above expression can be easily rewritten, taking $M= T_1$, as
\begin{equation}
\left.\Delta N_{\rm eff}\right\rvert_t / N_\nu \approx   4\pi\, g_{\rm DM} \int_0^\infty \frac{{\rm{d}}\uppi}{(2\pi)^3} \left(\frac{m \sqrt{\uppi}}{\mu_m}\right)^{4} \, \epsilon(\uppi, \mu_m)\, \left.f\left(\frac{\uppi}{\sqrt[3]{\varepsilon}}, \mu\right) \right\rvert_{t}/\rho_\nu \, ,
\end{equation}
where $\epsilon(\uppi, \mu_m)= \sqrt{\uppi^2+ \mu_m^2}-\mu_m$, $\mu_m= m/T$ and $\varepsilon= g_*^s(\mu)/g_*^s(\mu_1)$. 
As usual, $T$ is the photon temperature. The  process dependent amplitude $f_0$ is hidden inside $f(\uppi,\mu)$, which is the DM distribution function per degree of freedom, encapsulating the incomplete thermalization of the FIMPs. 

For example, the existence of a single DM species, fixing $\Omega_{\rm {DM}} = 0.26$ \cite{Ade:2015xua}, yields during the BBN period a 1$\sigma$ lower bound on the DM mass, namely
$
m \gtrsim 13\, \mathrm{eV} ,
$
less tight than the one inferred from the protohalo mass. 
This is a much more precise estimate of the deviation of the effective neutrino degrees of freedom than the one using a warm DM distribution for non-relativistic decays \cite{Kamada:2013sh,Strigari:2006jf}.
However, for the case of fermionic DM, the lowest allowed masses arise from the Tremaine-Gunn bound \cite{Boyarsky:2008ju}, which is tighter, $m \gtrsim 0.5$ keV. 
One notices that FIMPs at the keV-scale sourced by SM fields are compatible with BBN \cite{Cyburt:2015mya} and CMB \cite{Ade:2015xua} 1$\sigma$ measurements leading to $\Delta N_{\rm eff} \vert_{\rm BBN} \lesssim 0.01$. 
Moreover, assuming that the recombination takes place instantaneously at 0.3 eV, the DM modes are non-relativistic in this period, otherwise the Lyman-$\alpha$ bounds would be violated. 
This outcome could also explain the tension about the decrease of the deviation of the effective neutrino number from BBN to CMB-based measurements, such a difference $\Delta N_{\rm eff}\vert_{\rm CMB}- \Delta N_{\rm eff}\vert_{\rm BBN}<0$ can be delivered from the effective theory parameter space.

\subsection{Large-scale Cosmic Structure}

In this work, we assume that the dominant DM population is produced through processes involving SM modes. 
Furthermore, we show that such channels can be responsible for the present DM relic density. 

For example, we consider, without loss of generality, $1\rightarrow2$ decays of the Higgs modes into DM after the electroweak phase transition as production mechanism, and not Higgs scatterings at high energies. During this scenario, one should not consider annihilations, since they are suppressed in the phase space. 
Instead of  solving the Boltzmann equation approximately, following the methods discussed in \cite{Balducci:2017vwg}, we approach the problem in a more accurate way regarding the needs of this work. The natural way to compute the DM relic abundance is to evaluate the DM particle number at the present time $t_0$ using $n_{\text{DM}}= g_{\rm DM}\int \frac{{\rm{d}}^3p}{(2\pi)^3}f(p,T)\vert_{t_0}$. This yields
\begin{align}
&\frac{\Omega_{\rm{DM}} h^2}{0.12 } \approx \left(\frac{86.8}{g_*^s(\mu_1)}\right) \, \left(\frac{g}{2.3\times 10^{-9}}\right)^2\left( \frac{m}{{50\, \rm keV}} \right)\, ,\label{relic abundance}
\end{align} 
where  $g$ is the dimensionless feeble coupling between the Higgs field and the DM fields as introduced in generic interaction term \cref{interaction vertex}, taking $\Lambda = m_{\rm Higgs}$. This coupling appears in the distribution amplitude $f_0(\mu_1) = \mu_1^{-2} \frac{\lambda^2 m_h }{16\pi H(\mu_1)}$. 
This means that the DM modes should not freeze-in before the usual feeble freeze-in due to the Higgs decays; otherwise, the DM density would be reprocessed, destroying the FIMP assumption. 
Hence,  the mass of the DM particle is bounded from below, namely, we can infer $
m\gtrsim \mathcal{O}(10)\, \mathrm{eV},
$
which is less tight than the corresponding bound due to the cut-off masses of the previous section.
In the case of $2\rightarrow 2$ high energy DM production before the electroweak phase transition, a similar relation to \cref{relic abundance} appears after using the suitable phase-space distribution function \cref{phase space 2to2}.

In addition, since we are interested in purely feeble theories, the produced DM particles should not freeze-in via self-interactions, at least for processes of order $\mathcal{O}(\lambda^4)$, and $\mathcal{O}(\tilde{G}_F^2)$, for the scalar and fermionic DM, respectively; such processes could modify their abundance \cite{Hansen:2017rxr}. 
This strong constraint yields, for the case of the scalar field 
$
m \lesssim \mathcal{O}(10)\, \mathrm{keV}\,,
$
otherwise the $2\leftrightarrow4$ processes would modify $\Omega_{\rm{DM}}$, as noticed in \cite{Matsui:2015maa}, however, for fermionic DM, the bound is of order a few MeV due to the structure of the effective interaction. Therefore, we showed that the parameter space of feeble theories sourced via SM fields is sufficient to reproduce the the large-scale cosmic structure, keeping at the same time the FIMP assumption alive.

\section{The Parameter Space of Feeble Theories}
Finally, we discuss the parameter space of feeble theories sourced by SM fields, and how it accommodates complete solutions for the small-scale crisis of $\Lambda$CDM cosmology. 

The various constraints derived previously for the DM mass spectrum of the feebly interacting particles are either upper or lower bounds on the masses, since the (self-)couplings depend on the DM mass either due to the dwarf and cluster observations, or the relic density of the cold DM. 
One notices that the allowed subset of the parameter space, which also fulfils the requirements of the ETHOS-4 tuned model \cite{Vogelsberger:2015gpr}, and provides solutions at the same time to the enduring small-scale problems of the $\Lambda$CDM cosmology, is extremely narrow, at least in the case of scalar FIMPs. 
On the other hand, if we consider fermionic DM, the corresponding constraints differentiate themselves from the ones of the scalar case due to the Born-regime of the self-scatterings and the more involved thermal evolution towards the measured value of the DM relic abundance $\Omega_{\rm{CDM}}$. This results in less tight bounds regarding the mass spectrum and the couplings inside the dark sector, making the simultaneous solutions of the cusp vs. core, the too big to fail, the diversity and the missing satellite problems more viable and solid.

Now, we provide information about possible realizations presenting the most simple theories, which are still compatible with all present constraints. More precisely, we examine the existence of a minimal bosonic and fermionic extension of the SM sector. 
For simplicity, and without loss of of generality, we do not consider theories with mass-scales larger than the reheating temperature $T_0 \equiv T_{\text{RH}}$, where the SM modes are ultra-relativistic while producing the FIMPs.

The simplest paradigm of a single field theory is a singlet boson extension of the Higgs sector in accordance with the general operator \cref{interaction vertex}, setting $\Lambda = m_{\rm Higgs}$, together with the necessary dark self-interactions; namely, we assume the presence of an uncharged scalar field $X$ \cite{McDonald:2001vt}. Since the cluster scale constraints dictate $m\ll m_{\rm Higgs}$ for the case where the thermal production of the DM modes takes place after the electroweak phase-transition, the dominant channels are the feeble Higgs decays. In other words, the DM density is populated after the out-of-equilibrium decays of the Higgs particle. Schematically, the relevant part of the interaction Lagrangian density after the electroweak phase transition should take the form of

\begin{equation}
    \mathcal{L}_{\rm int}= g\, m_{\rm Higgs} h X^2 + \tfrac{\lambda}{4} X^4 \, .
\end{equation}
Here, $h$ denotes the Higgs field, and $g, \lambda$ perturbative couplings. Furthermore, in order to assume stability of DM structures, some symmetry should be imposed, e.g.\ a $\mathds{Z}_2$ symmetry of the singlet field. The self-interactions arise via the quartic self-coupling, denoted by $\lambda$.

We examine our parameter space and we find that a DM pure scalar field with mass of 3.55 keV and self-coupling of order $\mathcal{O}(10^{-7})$ realizes the ETHOS-4 tuned model, providing also solutions at the cluster scales \cite{Elbert:2016dbb} due to the constant $\mathcal{O}(0.1) \mathrm{cm}^2\mathrm{g}^{-1}$ SIDM cross-section. Even larger masses around 7-8 keV, which are allowed to exist in the parameter space of the scalar theory, are able to alleviate the small-scale crisis of $\Lambda$CDM.
Furthermore, at dwarf-scales, the $\Lambda$CDM anomalies are alleviated due to sufficiently large cut-off masses and the SIDM virtue of the theory. However, there exist indications that a velocity dependence of the self-interaction cross-sections should be present \cite{Tulin:2017ara}, making the single field theory unable to provide such results.  Moreover, such DM masses cannot be solely generated from the Higgs field and additional fine-tuning is necessary.

The absence of velocity-dependent SIDM cross-sections in the single field theory hints to a further extension of the dark sector by adding one more field, at least at low energies; namely, we assume that the FIMPs (bosons or fermions) are charged under a local symmetry with coupling $g'$ as in \cite{Karananas:2018goc}. Since the charged fields $F$ are FIMPs, there exists a sub-set of the parameter space where the dark force mediator $V$ does not even freeze-in. For example, the self-interaction Lagrangian density for fermionic DM takes the following form,

\begin{equation}
    \mathcal{L}_{\rm self}= g'\,\bar{F} \slashed{V} F\,.
\end{equation}

It is now clear that the velocity-dependence SIDM virtue of these theories arises naturally via dark forces from some 4-Fermi interaction, mediated by the hidden boson $V$ \cite{Tulin:2013teo}. This type of interaction imposes a further constraint on the parameters of the theory, namely, the SIDM scattering rate should not freeze-in. Otherwise, the DM relic density would be modified through DM annihilations to $V$'s, leading to a second WIMP miracle \cite{Bernal:2017kxu}. In addition, the masses are not free to choose, we demand that $m_V\lesssim m_F$ and, therefore, theories like the minimal scalar sourced sterile neutrinos \cite{Merle:2015oja,Konig:2016dzg} are unable to reproduce the SIDM paradigm. A part of the parameter space survives even after the application of this tight constraint leading to $m_V \sim \mathcal{O}(\rm eV)$ for $m_F \sim \mathcal{O}(\rm keV)$.

Moreover, the production of charged FIMPs could again take place via Higgs decays; namely we keep track of the correct dimensions and setting $\Lambda$ equal to $m_{\rm Higgs}$. The DM fields $\Phi_{\rm DM}$ could even couple to the Higgs field through higher-dimensional operators, as in \cite{Matsui:2015maa}, or even coupled to a SM matter fields. For example, the portal could be leptophilic, taking the generic term \cref{interaction vertex} for $\ell =5 $ or 6 for bosonic or femionic DM, respectively,

\begin{equation}
    \mathcal{L}_{\rm portal}= \frac{g}{\Lambda^{\ell-4}}\bar{\mu} \mu \, \Phi^2_{\rm DM}\, .
\end{equation}
Here, SM muons source the DM feebly production well before the electroweak phase transition. This operator could arise after integrating out heavier, non-dynamical fields. For a more detailed review of leptophilic interactions with fermionic DM, see \cite{Balducci:2018dms}. The production mechanism depends strongly on the reheating temperature, requiring that $T_0\equiv T_{\rm RH} < \Lambda$, as explained previously. 

We note that if the DM particles are sourced by the heavy DM scalar and not by the Higgs field directly, as in \cite{Balducci:2017vwg}, the corresponding allowed cut-off masses are almost $10\%$ smaller than those appearing in the single field theory, due to the larger values of $g_*^s$ in \cref{relic abundance} at the time of the DM production. The study of this case is beyond the scope of this work, since the DM modes are not sourced directly via SM fields.

\section{Discussion and Conclusion}

In this work, we explored the possibility of the indirect observation of feebly interacting dark matter through its signatures regarding the implied small-scale structure formation. 
We derived and studied its phase-space distribution functions, and showed that such particles, after obtaining their relic abundance and their interactions, are severely constrained if the generating feeble source are SM fields.  
We continued presenting a list of the most recent constraints on such theories, which arise from the cosmic small-scale structures of the proposed DM population.
The recent Lyman-$\alpha$ bounds \cite{Irsic:2017ixq}, together with the positive observations regarding the rotation curves, the missing dwarf satellites and the cored profiles of galaxies discussed in \cite{Vogelsberger:2015gpr, Schewtschenko:2015rno, Tulin:2017ara}, indicate the possible presence of new physics well below the weak scale: multi-keV DM masses and multi-eV masses for the force mediators. 
In other words, the feeble theories favor masses at the keV scale for the dark matter relic particles, i.e.\ a mass around $m \sim \mathcal{O}(\rm keV)$ is able to provide a solid solution to the enduring small-scale problems of $\Lambda$CDM, alleviating the missing satellite problem while solving the too big to fail, diversity and cusp vs. core problems. One can notice that possible annihilations, or even meta-stable decays, of such modes could correspond to the 3.55 keV mysterious photon-ray \cite{Bulbul:2014sua}.
More precisely, the derived protohalo masses alleviate the missing satellite problem and the SIDM cross-sections, together with the cut-off of the linear matter power spectrum, as explained in \cite{Vogelsberger:2015gpr}, and are able to flatten the cuspy profile and solve the massive subhalos issues of CDM simultaneously, and are compatible with all present constraints at the dwarf and cluster scales.
The result of this approach is even more interesting under the scope of the recent observations by the XENON collaboration \cite{Aprile:2020tmw}, finding an excess of the electron recoil energy below 7 keV, notably between 2-3 keV, as the focus on FIMPs alleviating the small-scale problems naturally leads to dark matter with masses in this region. 
It would be interesting to investigate the possible signals of these models in the XENON1T detector in more detail for specific examples.

The family of SM sourced theories could enable a minimalistic scenario of viable thermal evolution, solving the enduring small-scale challenges appearing in the $\Lambda$CDM cosmology. The proposed theories are  characterized by three checkpoints in the thermal evolution; in other words, three temperatures $T_0, \, T_1,\, T_2$, the reheating, the production, and the rest-mass temperatures respectively, are responsible for the resulting cosmic structure at large and small scales.

However, these toy models do not admit dark acoustic oscillations as ETHOS-4 tuned model, since no kinetic decoupling is present; therefore, further study and simulations are needed in order to examine the exact impact of the free-streaming cut-off on the dwarf scale. 
The ultra-relativistic limit of such theories could also be interesting to study, as long as new mass-scales above the reheating temperature are introduced.

\acknowledgments

It is a great pleasure to thank Stefan Hofmann for inspiring discussions.
This work was supported in part by the DFG cluster of excellence 'ORIGINS'.


\bibliographystyle{JHEP}
\bibliography{draftfim}

\end{document}